# Sausage instabilities on top of kinking lengthening current-carrying magnetic flux tubes

Jens von der Linden[a)] and Setthivoine You[b)]
*University of Washington, Seattle, Washington 98195-2400, USA*

We theoretically explore the possibility of sausage instabilities developing on top of a kink instability in lengthening current-carrying magnetic flux tubes. Observations indicate that the dynamics of magnetic flux tubes in our cosmos and terrestrial experiments can involve topological changes faster than time scales predicted by resistive magnetohydrodynamics. Recent laboratory experiments suggest that hierarchies of instabilities, such as kink and Rayleigh-Taylor, could be responsible for initiating fast topological changes by locally accessing two-fluid and kinetic regimes. Sausage instabilities can also provide this coupling mechanism between disparate scales. Flux tube experiments can be classified by the flux tube's evolution in a configuration space described by a normalized inverse aspect-ratio $\bar{k}$ and current-to-magnetic flux ratio $\bar{\lambda}$. A lengthening current-carrying magnetic flux tube traverses this $\bar{k}$–$\bar{\lambda}$ space and crosses stability boundaries. We derive a single general criterion for the onset of the sausage and kink instabilities in idealized magnetic flux tubes with core and skin currents. The criterion indicates a dependence of the stability boundaries on current profiles and shows overlapping kink and sausage unstable regions in the $\bar{k}$–$\bar{\lambda}$ space with two free parameters. Numerical investigation of the stability criterion reduces the number of free parameters to a single one that describes the current profile and confirms the overlapping sausage and kink unstable regions in $\bar{k}$–$\bar{\lambda}$ space. A lengthening, ideal current-carrying magnetic flux tube can therefore become sausage unstable after it becomes kink unstable. *Published by AIP Publishing.* [http://dx.doi.org/10.1063/1.4981231]

Magnetic flux tubes are defined as a bundle of magnetic field lines that run through a closed contour.[1] When current flows along a magnetic flux tube, several current-driven instabilities may develop, which are generally well understood with the ideal magnetohydrodynamic (MHD) behavior of current-carrying flux tubes,[2] where for mathematical convenience, the column is modeled by an infinitely long cylinder and classified into one of two ideal analytical models: a skin screw pinch (with an axial magnetic field and an electrical current flowing only on an infinitely thin boundary layer separating the plasma from vacuum)[3] or a diffuse pinch (with an axial magnetic field, no skin current, but some distribution of current inside the plasma).[4] In all cases, classical instability onset conditions are retrieved from the Energy Principle[5] on Fourier perturbed states of the idealized current-carrying magnetic flux tube, i.e., starting from a static equilibrium like a ball at rest on a hill or valley, if the perturbed potential energy is negative, the system would be unstable to that particular mode. Indications that the coupling of different types of instabilities may play a role in astrophysical, heliospheric, and laboratory phenomena have motivated a renewed interest in revisiting and extending classical instability criteria. MHD instabilities may be responsible for structure formation in astrophysical jets,[6] and understanding these structures may provide insight into activity at the source of the jets. Kink and sausage instabilities have been observed in coronal loops and could trigger large energy releases and heating of the solar corona.[7] Instabilities appear to be necessary for the formation of toroidal plasma configurations from initially open flux tubes, such as spheromaks[8,9] and spherical tori.[10] Recent experiments have identified an instability cascade, where a macroscopic instability triggers a microscopic instability, creating small scale structures which directly couple the disparate plasma scales:[11] the acceleration of a growing kink on the scale length of the flux tube acts as an effective gravity, leading to a Rayleigh-Taylor instability. The Rayleigh-Taylor instability forms structures on the scale of ion inertial lengths, enabling fast reconnection.[12] Two neighboring current-carrying magnetic flux tubes may kink on system scale lengths, attract each other, merge, reconnect ion inertial lengths, and launch ion-cyclotron waves.[11] Spheromak merging experiments observe annihilation of magnetic helicity followed by the formation of a Field-Reversed Configuration (FRC) with ion flows, when the scale lengths approach ion Larmor radii.[13,14] Partially toroidal flux tubes in a solar loop laboratory experiment erupt when they are unstable to both the torus and kink instabilities but fail to erupt when they are unstable to only one of the instabilities.[15] In toroidal fusion devices, the merging of discrete temperature flux tubes has been associated with global sawtooth events.[16] To identify possible couplings between MHD instabilities, it is necessary to examine current-carrying magnetic flux tube stability beyond the conditions used to derive classical stability criteria. Finite-length screw pinches with one or both ends tied down have kink instability criteria modified from classical Kruskal-Shafranov values,[17] explaining some experimental results.[18] Sheared axial plasma flow stabilizes the Z-pinch to kink and

---

[a)]Electronic mail: jensv@uw.edu
[b)]Electronic mail: syou@aa.washington.edu. URL: www.aa.washington.edu/research/youlab

sausage instabilities,[19] an effect subsequently observed in the laboratory.[20] In many flux tube experiments, the flux tube is bound between two electrodes and remains at a fixed length while the current ramps up. In contrast, a flared current-carrying magnetic flux tube generated from one boundary gives rise to strong axial plasma flows before collimating into a filamentary cylinder. This collimation occurs continuously as the flows convect magnetic flux into the flared end of the flux tube, resulting in a lengthening of the collimated flux tube while the current ramps up. This behavior was described theoretically[21] and observed experimentally.[22] The same flux tube has then been observed to become kink unstable, detach from the source electrodes, and form a spheromak.[8] The detachment of the flux tube may be initiated by a sausage-like instability, but comparisons between measurements and the theory currently rely on stability criteria derived for skin screw-pinch models of infinite length, even though the experiment's flux tube has a distributed current and lengthens during the shot duration. This paper presents a single general stability criterion for the onset of ideal MHD instabilities in a lengthening current-carrying magnetic flux tube with both diffuse internal and sharp skin currents and comparisons to numerical calculations.

Here, the general, minimized form of the perturbed potential energy of the system is subdivided into 3 terms: the first (internal) term is simplified with Newcomb's analysis[4] of internal stability; the second (skin) term is further simplified by considering Bellan's analysis[21] of a self-collimating flared flux tube; the third (vacuum) term is simplified by assuming that the wall is at infinity. Differing from standard treatments, none of the terms are subsequently set to zero in the analytical treatment.

The starting point is the general, minimized, 1D perturbed potential energy of the current-carrying magnetic flux tube $\delta W(\xi)$ which depends only on the radial displacement $\xi(r)$ of the flux surface at radial position r. This real 1D scalar expression is reduced from $\delta W(\vec{\xi})$, a complex expression of the infinitesimal 3D displacement $\vec{\xi} = \xi(r)\hat{r} + \eta(\phi)\hat{\phi} + \zeta(z)\hat{z}$, with classical assumptions: the idealized cylindrically symmetric flux tube with periodicity length L and radius a is surrounded by vacuum, no walls (wall at radius $r = b \to \infty$), and subdivided into the internal (plasma) volume between $0 \leq r \leq p$ and a thin shell (interface or skin) between $p \leq r \leq v$, where p is the "plasma" side of a and v is the "vacuum" side of a. The plasma pressure is isotropic and inviscid. The plasma is incompressible, $\nabla \cdot \vec{\xi} = 0$, which is the most unstable worst case scenario and allows reduction of the 3D $\vec{\xi}$ to a 1D $\xi$. This assumption also removes pressure-driven instabilities from consideration, leaving only the current-driven contributions. The fundamental instabilities are the sausage ($m = 0, n \geq 1$) and kink ($|m| \geq 1, n \geq 1$) modes, where m is the azimuthal mode number and n the longitudinal mode number of the Fourier perturbation on the cylinder. The perturbations are small and linear, to ignore higher order terms, and adiabatic, and so, the perturbed plasma pressure can be simply expressed in terms of the equilibrium pressure and the displacement $\xi$. The longest (worst-case, $n = 1$) perturbation wave length is assumed so that $k = 2\pi/L$. The system is described by ideal MHD, and so, the perturbed magnetic field can be expressed simply in terms of the plasma displacement and the equilibrium magnetic field. The system is assumed initially to be in static equilibrium, and so, equilibrium quantities are not time dependent and integration constants can be put to zero.

The linearly perturbed potential energy $\delta W$ is thus integrated by parts into contributions $\delta W_{pl}$ from the plasma, $\delta W_i$ from the interface, and $\delta W_v$ from the vacuum.[2] The $\delta W_{pl}$ term represents the available free energy of the plasma, with contributions from the internal vacuum magnetic field, the plasma pressure, and internal currents ($\delta W_{B_1}, \delta W_{p_1}, \delta W_{J_0}$, respectively). The $\delta W_i$ term represents the work required to move the boundary between the plasma and the vacuum. The $\delta W_v$ represents the energy transferred to the external vacuum magnetic fields. From the Energy Principle, the flux tube is stable if the total perturbed potential energy $\delta W = \delta W_{pl} + \delta W_i + \delta W_v \geq 0$ and unstable otherwise. The general, minimized form of the perturbed potential energy of the system is thus subdivided into three terms, and differing from standard treatments, none of the terms are subsequently set to zero in our analytical treatment. We simplify the first (internal) term with Newcomb's analysis[4] of internal stability, the second (skin) term with Bellan's analysis[21] of a self-collimating flared flux tube, and the third (vacuum) term with the assumption that the wall is at infinity. We now look at each term individually. The incompressible plasma's perturbed potential energy $\delta W_{pl}$ is given by[4]

$$\delta W_{pl} = \frac{\pi L}{2\mu_0} \left( \int_0^p \left[ f\left(\frac{\partial \xi}{\partial r}\right)^2 + g\xi^2 \right] dr + \left[ h\xi^2 \right]_0^p \right), \quad (1)$$

where $f = r(krB_z + mB_\phi)/(k^2r^2 + m^2), h = (k^2r^2B_z^2 - m^2B_\phi^2)/(k^2r^2 + m^2)$, and g is also a function of k, r, $B_z$, and $B_\phi$ that will soon be eliminated, and m is the azimuthal mode number. The first term of Eq. (1) can be minimized if $\xi$ is the solution to the associated Euler-Lagrangian equation

$$\frac{\partial}{\partial r}\left[ f \frac{\partial \xi}{\partial r} \right] - g\xi = 0. \quad (2)$$

Eq. (2) has a singular point in the flux tube whenever $f = 0$, i.e., wherever $q = m$, having defined $q = 2\pi r B_z/(LB_\phi)$, corresponding to rational mode surfaces $q = m/n$ with $n = 1$. Solutions $\xi$ thus only exist in regions between two singular points (called an "interval"). It is sufficient to only consider stability to the $m = 0, \pm 1$ modes because if a column is stable to both, it is also stable to higher order modes.[4] So, the "no singular point" condition becomes $q \neq 0, \pm 1$. Newcomb[4] proves extensively the conditions for the existence of solutions and the conditions for internal stability ($\delta W_{pl} \geq 0$) of a collimated magnetic flux tube: a collimated magnetic flux tube is stable if and only if (a) the Suydam criterion[23] is satisfied for $r > 0$ (but does not have to be at $r = 0$); (b) the Euler-Lagrangian solutions satisfy the "small" conditions on the left-hand side of an interval, namely,

$$\xi(0) = 0 \quad \text{if} \quad m \neq \pm 1 \quad \text{and} \quad \frac{\partial \xi}{\partial r} = 0 \quad \text{if} \quad m = \pm 1 \quad (3)$$

and (c) the Euler-Lagrangian solution does not vanish twice in any interval. The existence of singularities and solutions are dependent on the experimental conditions, and so, numerical codes are generally employed to solve Eq. (2). But we proceed analytically for now.

The lengthening flux tube is initially stable; therefore, the three conditions (a)–(c) are fulfilled. Early in the discharge, the collimated flux tube is short ($L$ small) and stable (small $B_\phi$), and so, $q \approx rB_z/(LB_\phi) \gg 1$. As the collimated flux tube lengthens and current increases, $q$ approaches one from above and is always non-zero. Hence, just before the onset of instability, there are no singular points inside the flux tube and we can consider the whole plasma cross-section as a unique "interval" where the three Newcomb stability conditions are fulfilled, in particular, conditions (b) and (c). So, except at $r = 0$, the solution $\xi(r) \neq 0$ and the Euler-Lagrangian equation (2) is satisfied, giving $g = \partial/\partial r[f \partial \xi/\partial r]/\xi$, which can be substituted into Eq. (1) to give

$$\delta W_{pl} = \frac{\pi L}{2\mu_0} \left( \left[ f\xi \frac{\partial \xi}{\partial r} \right]_0^p + \left[ h\xi^2 \right]_0^p \right). \quad (4)$$

Strictly, the lower bound is at $r_c < a$ with $r_c \to 0$, but those terms vanish since, at the flux tube center, the solutions all satisfy the Newcomb small-value conditions $\xi(0) = 0$ or $\xi'(0) = 0$ where ' denotes $\partial/\partial r$. At the plasma edge $r = p$, we take $\xi(p) = \xi(a) \equiv \xi_a$, $\xi'(p) \approx \xi_a \delta/a$, where $\delta$ represents a "rigidity" of the displacement of the boundary and $f(p) = f(a)$. The plasma term Eq. (4) thus becomes

$$\delta W_{pl} = \frac{\pi L}{2\mu_0} \delta \frac{f(a)}{a} \xi_a^2 + \frac{\pi L}{2\mu_0} h(a) \xi_a^2. \quad (5)$$

For the interface term $\delta W_i$, the standard derivation[24] of the perturbed potential energy gives

$$\delta W_i = \frac{\pi a L}{2} \xi_a^2 \left( \left[ \frac{\partial}{\partial r} \frac{B^2}{2\mu_0} \right]_v - \left[ \frac{\partial}{\partial r} \left( P + \frac{B^2}{2\mu_0} \right) \right]_p \right), \quad (6)$$

where $B^2 = B_\phi^2 + B_z^2$ and the plasma pressures $P$ are the equilibrium values. This term only exists if a surface skin current exists and represents the work required to move the plasma boundary. Since we assumed the starting point to be a static, axisymmetric equilibrium, then the pinch force $-B_\phi/(\mu_0 r)$ balances the total pressure gradient $\partial/\partial r[P + B^2/(2\mu_0)]$. The "jump" in the current profile from the core plasma current to the skin current value is represented by $B_\phi(p)$ being different from $B_\phi(v)$, giving

$$\delta W_i = \frac{\pi L}{2\mu_0} \xi_a^2 \left( B_{\phi p}^2 - B_{\phi v}^2 \right). \quad (7)$$

The classical skin current model assumes no plasma current inside the flux tube, i.e., $B_{\phi p} = 0$, but we keep both terms here. The vacuum contribution $\delta W_v$ is given by the usual expression[24] (with the wall radius $b \to \infty$)

$$\delta W_v = -\frac{\pi L}{2\mu_0} (kaB_{zv} + mB_{\phi v})^2 \frac{\xi_a^2}{ka} \frac{K_m(|ka|)}{K_m'(|ka|)}, \quad (8)$$

where $K_m$ is the modified Bessel function of the second kind of order $m$. Reassembling Eqs. (5), (7), and (8) gives the total perturbed potential energy as

$$\begin{aligned}\delta W =\ & \frac{\pi L}{2\mu_0} \delta \frac{f(a)}{a} \xi_a^2 + \frac{\pi L}{2\mu_0} h(a) \xi_a^2 \\ & + \frac{\pi L}{2\mu_0} \xi_a^2 \left( B_{\phi p}^2 - B_{\phi v}^2 \right) \\ & - \frac{\pi L}{2\mu_0} (kaB_{zv} + mB_{\phi v})^2 \frac{\xi_a^2}{ka} \frac{K_m(|ka|)}{K_m'(|ka|)},\end{aligned} \quad (9)$$

which is compared to zero to determine the onset of stability. After dividing by $\pi L/(2\mu_0)\xi_a^2$ and defining $\bar{k} \equiv ka$ and $m \to -m$, for convenience since for the kink, the most unstable modes will be perpendicular to the magnetic field ($kB_\theta + mB_z = 0$), the flux tube is stable if

$$\begin{aligned}& \frac{(\delta+1)\bar{k}^2 B_{zp}^2 + (\delta-1)m^2 B_\phi^2 + 2\delta m\bar{k} B_{\phi p} B_{zp}}{\bar{k}^2 + m^2} + B_{\phi p}^2 - B_{\phi a}^2 \\ & - \frac{(mB_{\phi v} - \bar{k}B_{zv})^2}{\bar{k}} \frac{K_m(|\bar{k}|)}{K_m'(|\bar{k}|)} > 0.\end{aligned} \quad (10)$$

This is normal as far as we can go since $B_{zp}$, $B_{za}$, $B_{zv}$ and $B_{\phi p}$, $B_{\phi a}$, $B_{\phi v}$ are generally independent parameters that have to be determined experimentally. To continue, we remember $\lambda = \frac{\mu_0 I}{\psi}$ with $\psi = B_z \pi r^2$, and so, Ampere's law $B_\phi = \mu_0 I/(2\pi r)$ can be rewritten as

$$\lambda_p = \frac{2B_{\phi p}}{aB_{zp}} \quad \text{and} \quad \lambda_v = \frac{2B_{\phi v}}{aB_{zv}} \quad (11)$$

for the plasma side and the vacuum side, respectively. Previous experimental results[22] showed our starting point (collimated magnetic flux tube) results from MHD pumping in a flared flux tube.[21] The MHD pumping theory points out that once collimated, the magnetic flux tube must have a uniform axial magnetic field across the interface, i.e., $B_{zp} = B_{za} = B_{zv}$, where the plasma current is entirely axial. Therefore, if the flux tube is fully collimated at the onset of instability, after defining dimensionless $\bar{\lambda} \equiv \lambda_v a$ and $\lambda_p a \equiv \epsilon\bar{\lambda}$, the stability condition Eq. (10) reduces to

$$\begin{aligned}& \frac{[2\bar{k} - m\epsilon\bar{\lambda}][(\delta+1)2\bar{k} - (\delta-1)m\epsilon\bar{\lambda}]}{\bar{k}^2 + m^2} \\ & + (\epsilon^2 - 1)\bar{\lambda}^2 - \frac{(m\bar{\lambda} - 2\bar{k})^2}{\bar{k}} \frac{K_m(|\bar{k}|)}{K_m'(|\bar{k}|)} > 0\end{aligned} \quad (12)$$

with the conditions $B_{\phi v} = B_{\phi a} \neq 0$ and $\bar{\lambda} \neq 0$. If the left-hand side of (12) is negative for $m = 0$, then the flux tube is sausage unstable and if it is negative for $m = 1$, it is kink unstable. The Kruskal-Shafranov $m = 1$ kink condition can be retrieved from (12) by bearing in mind that it is usually derived for a long, thin column ($k \to 0$) with a skin current and no internal currents ($\epsilon = 0$). For small argument, $K_1(x) \to x^{-1}$ and $K_1'(x) \to -x^{-2}$, which simplifies (12) to the Kruskal-Shafranov condition $\bar{k} > \bar{\lambda}/2$ for $\delta \to 0$. Remembering that $\bar{k} = 2\pi a/L$ and $\bar{\lambda} = \mu_0 Ia/\psi$, where $\psi = B_z \pi a^2$ and $\mu_0 I = B_\phi/(2\pi a)$, gives the well-known $2\pi a B_z/(LB_\phi)$ in cylindrical geometry. The Tayler criterion[25] for sausage $m=0$ stability in a skin screw-pinch can also be retrieved from (12) by taking the same limits. One obtains $\bar{\lambda} > 2\sqrt{2}$ for $\delta \to 1$, equivalent to the more familiar $B_z^2 > B_\phi^2/2$.

The stability condition (12) thus maps out a stability space (Fig. 1). The kink instability region is hatched, to the right of the curve that approaches the classical Kruskal-Shafranov condition (dotted line $\bar{k} = \bar{\lambda}/2$). The region that is both sausage and kink unstable is cross-hatched, to the right of the curve that falls on the $\bar{k} = 0$ axis away from the origin. A short, low current magnetic flux tube ($\bar{k} \to$ large, $\bar{\lambda} \to 0$) thus starts to become stable at the top left of the space (white region). As the column lengthens and current increases ($\bar{k} \to 0$, $\bar{\lambda} \to$ large), the column's ($\bar{k}$; $\bar{\lambda}$) point travels to the bottom-right of the space, crossing the $m=1$ threshold to be kink unstable (hatched region). Eventually, the column may cross the $m=0$ threshold and becomes sausage unstable (cross hatched). The parameter $\epsilon = \lambda_p a/\bar{\lambda}$ represents the ratio of the internal current to the skin current and characterizes the "hollowness" of the current profile across the flux tube, e.g., if $\epsilon = 0$, the plasma column has no internal current, only a skin current. The parameter $\delta = a\xi'(a)/\xi(a)$ represents the "abruptness" or "rigidness" of the displacement of the plasma boundary with respect to the internal plasma. If $\delta = 0$, the plasma cross-section moves rigidly radially without compressing, as in a kink instability. If $\delta = 1$, then the boundary layer displacement is greater than the internal layers' displacement, as in the compressing sausage instability. Eq. (12) thus demonstrates three things: (1) a dependence of both the kink and sausage instability boundaries on the current profile $\epsilon$; (2) a significant region that is both kink and sausage unstable; and (3) a dependence of the instability boundaries on the value of the rigidness parameter $\delta$.

Although we cannot solve analytically for $\delta$, its value is uniquely determined for given periodicity numbers ($m$ and $\bar{k}$), axial magnetic field ($\bar{\lambda}$), and current profile ($\epsilon$) by the Euler-Lagrange equation (Eq. (2)). Numerical integration of Eq. (2) will thus constrain our value of $\delta$. Eq. (2) is a second order ordinary differential equation that can have several singularities. For numerical integration, the skin current region can no longer be assumed to be infinitesimal, and so, we construct profiles with a skin region of finite thickness. To continue, the equilibrium quantities will be expressed as dimensionless quantities (denoted by an over bar). The dimensionless axial current density $\bar{j}_z$ is described by a step function (Fig. 2); in the core, the current density is constant, rises smoothly to the skin current, and drops off smoothly to zero at the plasma edge. The smoothly varying transition regions are described by fourth order polynomials, continuous with the core and skin regions up to the second derivative. Since $\bar{j}_z$ smoothly varies to zero, the parameter $\epsilon = 0$. The current profile is described by the fraction of core-to-total current $\epsilon_{eff} = I_{core}/I_{total}$, where $I_{core}$ is the current in the constant current core region and $I_{total}$ is the total current driven along the flux tube. The azimuthal magnetic field $\bar{B}_\theta$ is determined by Ampere's law. The axial magnetic field $\bar{B}_z$ is taken to be constant across the flux tube. The pressure gradient $\bar{p}'$ is balanced by $\bar{j}_z \bar{B}_\theta$. Given the above assumptions, all profiles are uniquely determined by the dimensionless numbers $\epsilon_{eff}$, $\bar{k}$, and $\bar{\lambda}$. A fourth dependent dimensionless number is the ratio of plasma pressure to magnetic pressure at the axis $\beta_0$.

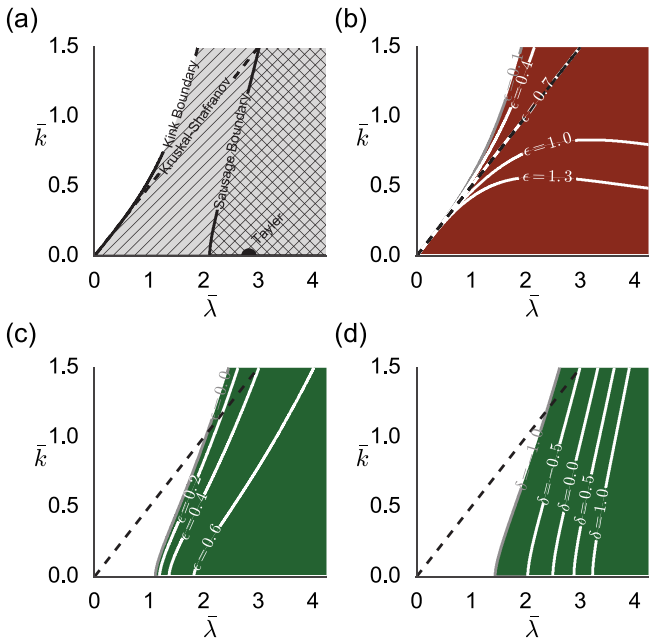

FIG. 1. Analytical $\bar{k}$–$\bar{\lambda}$ stability spaces parameterized with core current fraction $\epsilon$ and rigidness $\delta$. (a) Stability spaces for $\epsilon = 0.1$ and $\delta = 0.1$. The white region is stable, the gray region is unstable, the hatched region is kink ($m=1$) unstable, and the crosshatched region is unstable to both kink and sausage ($m=0$) modes. (b) Kink stability space dependence on $\epsilon$ with $\delta = 0$. The dark red region is unstable. (c) Sausage stability space dependence on $\epsilon$ with $\delta = 0.7$. The dark green region is unstable. (d) Sausage stability boundary dependence on $\delta$ with $\epsilon = 0.2$. The dark green region is unstable.

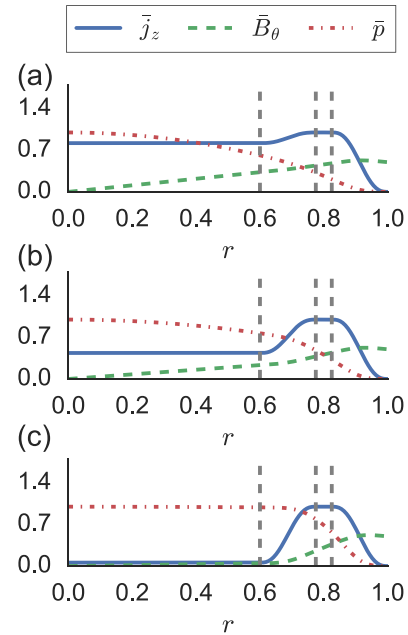

FIG. 2. Radial Profiles of $\bar{j}_z$, $\bar{B}_\theta$, and $\bar{p}$ for (a) $\epsilon_{eff} = 0.7$, (b) $\epsilon_{eff} = 0.5$, and (c) $\epsilon = \epsilon_{eff} = 0.1$ used for the calculations shown in Fig. 3. Dashed vertical lines demarcate (from left to right) core, transition, skin, and transition regions.

Since the Euler-Lagrange equation may have several singularities, the adaptive solver LSODA from ODEPACK is used. The solver is accessed through wrappers in the Python scipy.integrate library.[26] The Euler-Lagrange equation is re-expressed as a system of two first order ordinary differential equations (ODEs)

$$\frac{d\bar{\xi}}{d\bar{r}} = \bar{\xi}' \quad \text{and} \quad \frac{d}{d\bar{r}}\left(\bar{f}\bar{\xi}'\right) = \bar{g}\bar{\xi}. \quad (13)$$

LSODA is an adaptive solver, where the stepsize is modified to achieve the desired accuracy. So, a request for the value of the derivatives can be made at any given $r$. To accommodate these requests, the profiles are stored as cubic splines created with FITPACK routines accessed through Python wrappers in scipy.interpolate. Each interval, bounded by the singularities, can be integrated separately using the small solution Eq. (3) as an initial value to the right of the singularity. Since this study is only concerned with external stability, it will suffice to test the equilibrium for Suydam stability at the singularities and integrate only the last interval.

To determine the initial value to the left of the singularity, we use the Frobenius power series method for regular singular points.[27] The Frobenius power series expansion gives the relationship between $\bar{\xi}$ and $\bar{\xi}'$ close to the singularity at $\bar{r} = 0$, for $m = 0$ ($\bar{\xi} \sim C\bar{r}^1$ and $\bar{\xi}' \sim C\bar{r}^0$) and for $m \neq 0$ ($\bar{\xi} \sim C\bar{r}^{|m-1|}$ and $\bar{\xi}' \sim C|m-1|\bar{r}^{|m-2|}$). Since the constant $C$ cannot be determined, the magnitude of $\bar{\xi}$ is arbitrary and only the ratio $\bar{\xi}'/\bar{\xi}$ has meaning. At the non-geometric ($\bar{r} \neq 0$) singularities, the Frobenius method gives a quadratic indicial equation

$$n_\pm = -\frac{1}{2} \pm \sqrt{\frac{1}{4} + \frac{\beta}{\alpha}} \quad (14)$$

with

$$\alpha = \frac{\bar{r}_s \bar{B}_\theta^2 \bar{B}_z^2}{\bar{B}^2}\left(\frac{\bar{q}'}{\bar{q}}\right)^2 \quad \text{and} \quad \beta = \frac{2\bar{B}_\theta^2 \beta_0}{\bar{B}^2}\frac{d\bar{p}}{d\bar{r}}, \quad (15)$$

where $\bar{r}_s$ is the radial position of the singularity. The initial values for $\bar{\xi}$ to the right of the singularity are given by $|\bar{r} - \bar{r}_s|^{n_\pm}$. The solution with the larger $n$ is the "small" solution.[4] The small solution may diverge at $\bar{r}_s$ but will be well behaved away from the singularity. If $\frac{\beta}{\alpha} < -\frac{1}{4}$, the exponents $n$ will be complex, resulting in rapid oscillations through $\bar{\xi} = 0$ around the singularity. The condition for complex exponents is equivalent to the Suydam instability.[27] These initial values are used to integrate $\bar{\xi}$ for the given magnetic field profiles in the last interval of the Euler-Lagrange equation. The values of $\delta$ and $\bar{\xi}_a$ are then substituted into Eq. (12) with $\epsilon$ set to zero, to determine the $\delta W$ of each mode.

The perturbed potential energy $\delta W$ is calculated at $50 \times 50$ points in the $\bar{k}$–$\bar{\lambda}$ space and then interpolated throughout the space,[28] Zenodo. http://dx.doi.org/10.5281/zenodo.230489. The results of numerical calculations of the $\bar{k}$–$\bar{\lambda}$ stability space are shown in Fig. 3 for three values of the current fraction $\epsilon_{eff}$: 0.7, 0.5, and 0.1. In contrast to the analytical approach in Fig. 1, the numerical integration shows that the rigidness parameter $\delta$ is not a constant but varies with $\bar{k}$ and $\bar{\lambda}$ although the quantitative behavior is similar. The magnitude of $\delta W$ is arbitrary in the variational approach; however, we can compare relative magnitudes (all $\delta W$ values plotted are normalized). The kink unstable region matches the Kruskal-Shafranov condition for the long-thin regime, low $\bar{k}$ (Figs. 3(a)–3(c)). In regions of high $\bar{\lambda}$, $\beta_0$ becomes large and the $\beta$ term in Eq. (14) dominates. Since the pressure gradient for the chosen current profiles is always negative, any singularity due to the safety factor $q$ crossing a rational mode will result in infinite oscillations of the displacement $\xi$, i.e., be Suydam unstable. Eq. (2) is the $\rho\omega^2 \to 0$ case of the general cylindrical eigenvalue problem, where $\rho$ is the plasma density, $\omega$ is the mode frequency, and $\omega^2$ is the eigenvalue. The solutions to the eigenvalue problem are Sturmian, and their zero-crossings decrease with decreasing values of the eigenvalue $\omega^2$.[27] This means that Suydam-unstable profiles are likely unstable to the external kinks, as decreasing zero-crossings allow the eigenfunction $\xi$ to match the vacuum boundary conditions. To express this likely external kink instability, without solving the full eigenvalue problem, the Suydam-unstable regions are cross-hatched in Figs. 3(a)–3(c). The stability boundaries of both the sausage mode and kink modes depend on the core-to-total current fraction $\epsilon_{eff}$ (Figs. 3(d)–3(f)). As in the analytical results, there is a significant region which is unstable to both the kink and sausage modes. For high $\epsilon$ and high $\bar{\lambda}$, the solution to the Euler-Lagrange equation indicates external stability; however, $\xi$ does cross zero in these cases, suggesting internal instability. Again, taking the Sturmian property of the general eigenvalue problem into account, these zero crossings will likely allow the eigenfunction $\xi$ to match vacuum boundary conditions at lower $\omega^2$. To express this likely sausage instability, the internally unstable region is cross-hatched in Fig. 3(d). While the variational approach cannot determine the linear growth rates, the ratio of the potential energies $\delta W_{m=0}/\delta W_{m=-1}$ determines which mode will have the faster growth rate. In the region where both sausage and kink instabilities occur, the sausage instability has a larger linear growth rate for higher $\bar{k}$ values (shorter, fatter tubes), and this boundary shifts downward as the $\epsilon_{eff}$ decreases (Figs. 4(a)–4(c)). This suggests that a sausage could develop rapidly on top of a slowly growing kink.

The analytical and numerical stability conditions present evidence for the possibility of a sausage instability developing on top and after a kink instability in a lengthening current-carrying magnetic flux tube, particularly in the general case of a flux tube with internal *and* skin currents. A sausage instability in a screw pinch plasma may be unfamiliar because classical analyses consider diffuse or skin currents separately; however, evidence for sausage instabilities has been observed in coronal loops[30] and predicted in magnetospheric current sheets[31] and cylindrical liner compression experiments with axial fields.[32] We note that in ideal MHD, a linear configuration without magnetic field reversal is never sausage unstable without also being unstable to the kink. Sausage instabilities are commonly observed in Z-pinches,[33] which are always unstable to the kink. Likewise, for current-carrying magnetic flux tubes, the sausage unstable regions are inside the kink unstable regions of the $\bar{k}-\bar{\lambda}$ space. In magnetic flux tubes with helical fields, kinks have been observed to redistribute the current path and amplify the axial field.[8] The increased axial field could stabilize both the kink and sausage mode. A sausage instability is thus most likely to occur if the magnetic flux tube can quickly evolve to the upper right quadrant of the $\bar{k}-\bar{\lambda}$ configuration space where the sausage mode growth rates dominate. The linear growth rates will be on the order of the Alfvén transit time; however, nonlinear effects may lead to saturation of the kink instability as is observed in other experiments.[34] A sausage instability has been observed forming on top of a kinking current-carrying liquid mercury column.[35]

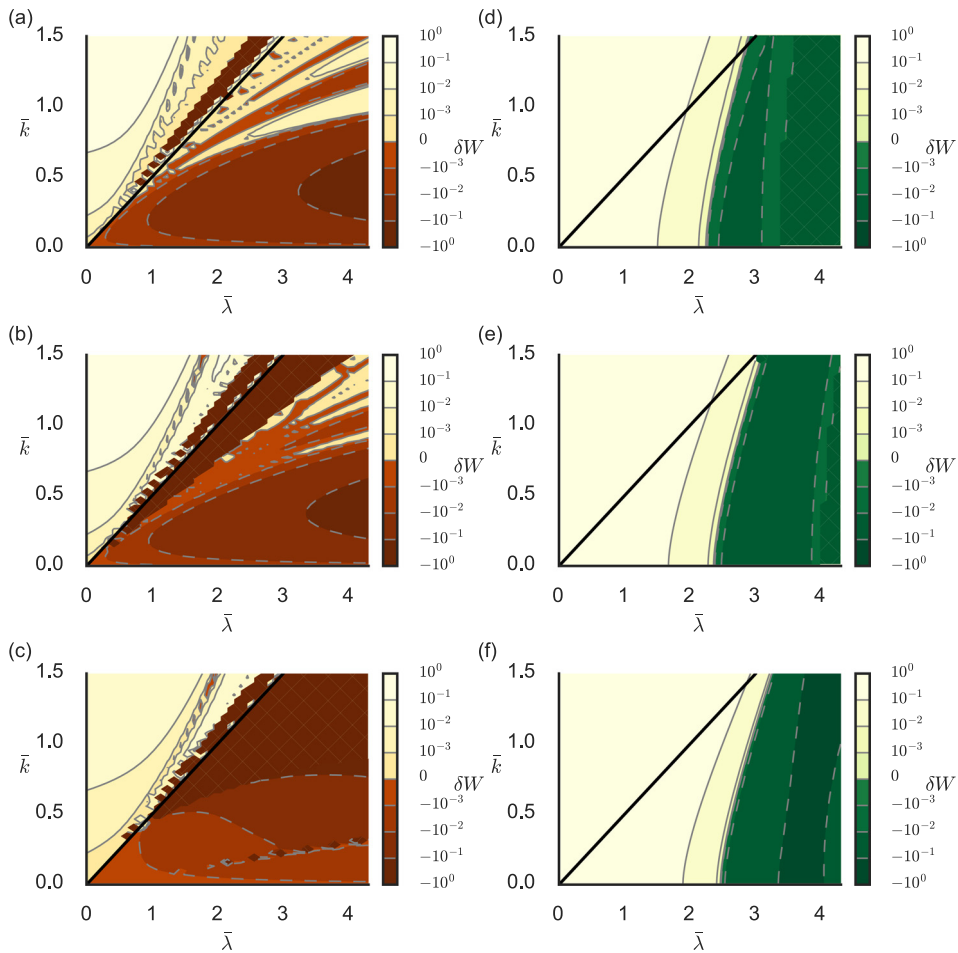

FIG. 3. Numerical $\bar{k} - \bar{\lambda}$ stability spaces with relative growth rates. Normalized kink $\delta W_{m=1}$ contours for (a) $\epsilon_{eff} = 0.7$, (b) $\epsilon_{eff} = 0.5$, and (c) $\epsilon_{eff} = 0.1$; Normalized sausage $\delta W_{m=0}$ contours for (d) $\epsilon_{eff} = 0.7$, (e) $\epsilon_{eff} = 0.5$, and (f) $\epsilon_{eff} = 0.1$. The cross hatched regions indicate Suydam unstable regions in the $m=1$ plots and regions with internal instabilities in the $m=0$ plots. (Associated dataset available at http://dx.doi.org/10.5281/zenodo.230611).[29]

Another crucial feature of the $\bar{k}-\bar{\lambda}$ stability space is the dependence of both the kink and sausage instability on the current distribution $\epsilon$. This effect is not observed in linear plasma experiments driven by washer guns[18,36] since the current ramp-up occurs on a much longer time scale than the lengthening of the flux tube. Washer-gun driven flux tubes are confined to low $\bar{k}$ when appreciable $\bar{\lambda}$ values are reached (bottom right of $\bar{k}-\bar{\lambda}$ space). In the long-thin (low $\bar{k}$) regime, the kink instability boundary always matches the classical Kruskal-Shafranov condition (assuming ideal periodic boundaries). Plasma jets produced by planar plasma gun experiments, however, evolve in $\bar{k}$ and $\bar{\lambda}$ on the same time scale. In these plasma jets, rapid pinching of the plasma from the gun followed by detachment has been observed at high $\bar{\lambda}$, which may be an indication for a sausage instability.[8] Solar loop stability experiments have noted the effect of finite aspect ratios (large $\bar{k}$).[15]

In regard to experiments, it is important to understand what implications a kink-sausage instability might have and if there are any unique signatures. In Z-pinch discharges, sausage instabilities are associated with beams of high energy ions.[33,37] The ions could be accelerated by a strong electric field generated from the increased resistivity in the pinched region of the sausage mode. While the magnetic field in a Z-pinch is purely azimuthal, in a current-carrying magnetic flux tube, the field is helical with shear across the magnetic flux surfaces. The compression of these sheared magnetic flux surfaces would provide a favorable magnetic topology for reconnection. An interesting question is whether a kink sausage instability could couple the MHD system to microscopic scales of ion inertial lengths or ion Larmor radii where two-fluid and kinetic effects become important and can lead to fast reconnection.[12] The assumptions of MHD are invalid when the ratio of drift over Alfvén velocity is close to or exceeds unity. At high drift velocities, kinetic effects such as wave-particle interactions and the decoupling of perpendicular ion and electron motion described by the Hall term could dominate. To understand when this may occur, it is helpful to express the ratio of the average drift velocity $v_d = I_z/(\pi a^2 n q)$ and the Alfvén velocity $v_A = B/\sqrt{\mu_0 n_0 m_i}$ in terms of the configuration space parameters, where $n_0$ is the number density, $q$ the charge, $m_i$ the ion mass, and $\mu_0$ the magnetic permeability. Starting with the velocity definitions

$$\frac{v_d}{v_A} = \frac{I_z}{\pi a^2 n_0 q} \frac{\sqrt{\mu_0 n_0 m_i}}{\sqrt{B_{\theta v}^2 + B_{zv}^2}}, \quad (16)$$

we can cancel a power of the azimuthal magnetic field after replacing the current with Ampère's law $I_z = 2\pi a B_\theta/\mu_0$

$$\frac{v_d}{v_A} = \frac{\sqrt{m_i}}{\sqrt{\mu_0 n_0 q^2} a} \frac{2}{\sqrt{1 + \frac{B_{zv}^2}{B_{\theta v}^2}}}. \quad (17)$$

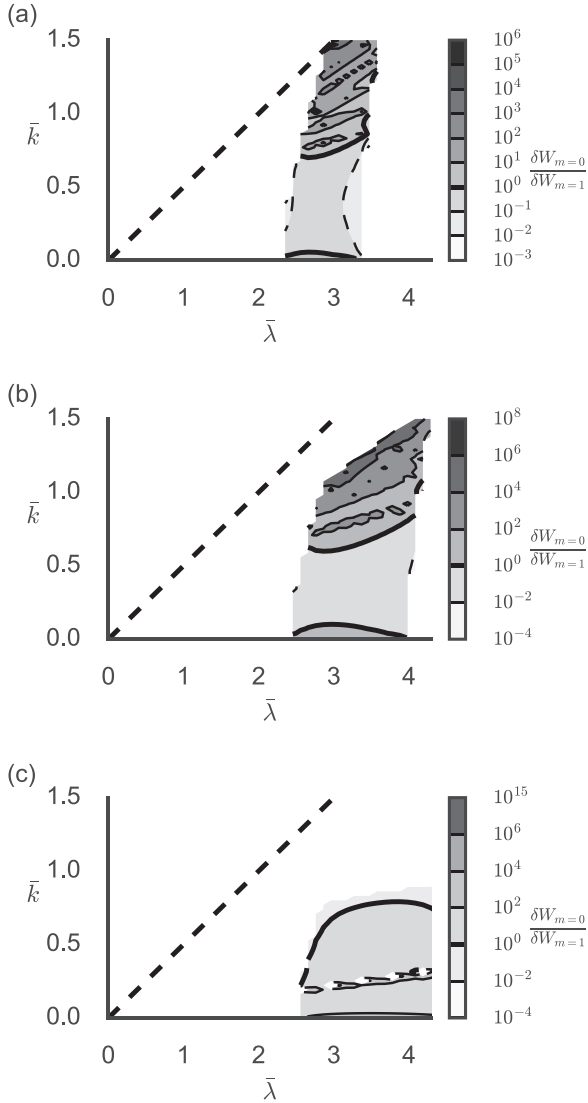

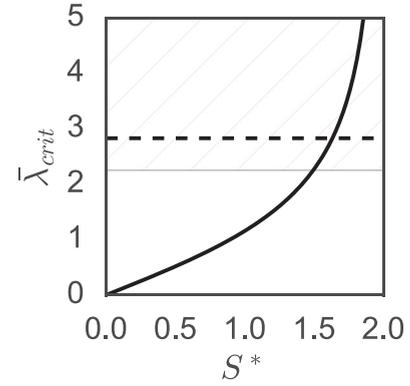

FIG. 4. The ratio of $\delta W_{m=0}/\delta W_{m=1}$ for (a) $\epsilon_{eff}=0.7$, (b) $\epsilon_{eff}=0.5$, and (c) $\epsilon_{eff}=0.1$. In the white space, either at least one of the potential energies is positive or undetermined. (Associated dataset available at http://dx.doi.org/10.5281/zenodo.230611).[29]

Substituting the magnetic field ratio $\bar{\lambda}_v$ Eq. (11) and the ratio of the characteristic length over the ion skin depth $\lambda_i$, also called the size parameter $S^* = a/\lambda_i = a\sqrt{\mu_o n_0 q_i^2/m_i}$, choosing the radius $a$ as the characteristic length, and assuming singly ionized plasmas, we obtain

FIG. 5. Critical $\bar{\lambda}$ value for a given size parameter $S^*$ above which the instability will couple to microscopic scales. The hatched region starts at the minimum sausage unstable $\bar{\lambda}$ for an $\epsilon_{eff}=0.7$ current profile. The dashed line denotes the Tayler criterion $\bar{\lambda} = 2\sqrt{2}$.

$$\frac{v_d}{v_A} = \frac{1}{S^*} \frac{2}{\sqrt{1 + \frac{4}{\bar{\lambda}_v^2}}}. \quad (18)$$

The critical value of $\bar{\lambda}_v$ at which the velocity ratio becomes unity is given by

$$\bar{\lambda}_{vcrit} = \frac{2S^*}{\sqrt{4 - S^{*2}}}. \quad (19)$$

This condition has a singularity at $S^* = 2$ with a relative shallow slope before $S^*1.5$, indicating that for low $S^*$, a sausage instability could couple to microscopic scales (Fig. 5). Low values of $S^*$ can be achieved with plasmas with higher ion masses and lower densities and by reaching the top right corner of the $\bar{k}-\bar{\lambda}$ configuration space where the flux tube radius is still large. Sausage instabilities can still occur when MHD dominates, at high $S^*$, without coupling to kinetic and two-fluid scales.

This work presents a general stability condition for current-carrying magnetic flux tubes with a wide range of aspect ratios, current-to-magnetic flux ratios, and current profiles. The analytical and numerical results point to sausage instabilities on top of kinking flux tubes as a possible cascade of instabilities. Further numerical and experimental studies are in progress to verify the $\bar{k}-\bar{\lambda}$ configuration space and determine growth rates of the kink and sausage instabilities. A new triple electrode planar plasma gun, Mochi.Labjet, is designed to generate magnetic flux tubes with discrete core and skin currents evolving over a wide range of the $\bar{k}-\bar{\lambda}$ stability space. Preliminary results indicate that the magnetic flux tube life time is several Alfvén transit times, while the peak rate of increase of $\bar{\lambda}$ is at least an order of magnitude faster than the Alfvén transit time and four times faster than the rate of increase of $\bar{k}$. This indicates that the sausage unstable region of the $\bar{k}-\bar{\lambda}$ configuration space should be accessible. The kink-sausage instability cascade could couple to two-fluid and kinetic regimes when the size parameter $S^*$ is small. This can be achieved with high ion masses, e.g., argon or krypton plasmas. High spatial resolution magnetic probe arrays will identify the cascade of MHD instabilities.

This material was based on work supported in part by U.S. DOE Grant No. DE-SC0010340 and the U.S. Department of Energy, Office of Science, Office of Workforce Development for Teachers and Scientists, Office of Science Graduate Student Research (SCGSR) Program. The SCGSR Program is administered by the Oak Ridge Institute for Science and Education (ORISE) for the DOE. ORISE is managed by ORAU under Contract No. DE-AC0506OR23100. S. You acknowledges P. M. Bellan for initiating discussion on sausage and kink instabilities.